\pdfoutput=1

\input{aipcheck}

\documentclass[
    ,final            
  ]
  {aipproc}

\layoutstyle{8x11single}

\def\lesssim{\mathrel{\mathpalette\vereq<}}

\makeatletter
\def\vereq#1#2{\lower3pt\vbox{\baselineskip1.5pt \lineskip1.5pt
\ialign{$\m@th#1\hfill##\hfil$\crcr#2\crcr\sim\crcr}}}
\makeatother

\begin{document}

\title{Looking up at seesaw and GUT scales from TeV}

\classification{14.60.Pq,12.60.Jv,12.10.-g,12.10.Kt,23.40.-s,95.35.+d,98.80.Cq}
\keywords      {neutrino, unification, double beta decay, dark matter,
  inflation, gravitational wave, accelerator, supersymmetry}

\author{Hitoshi Murayama}{
  address={Institute for the Physics and Mathematics of the Universe,
    University of Tokyo, Chiba 277-8568, Japan}
  ,altaddress={Berkeley Center for Theoretical Physics, University of
    California, Berkeley, CA 94720, USA}
  ,altaddress={Theoretical Physics Group, Lawrence Berkeley National
    Laboratory, MS 50A-5104, Berkeley, CA 94720, USA}
}

\begin{abstract}
  In this talk, I discuss how we may approach physics at the seesaw-
  and GUT-scales using data from the TeV scale.  Even though we cannot
  hope to directly reach these energy scales using particle
  accelerators, we may get convinced of grand unification and seesaw
  mechanism based on experimental data if Nature is kind to us.  In
  addition, we may find compelling reason to believe in leptogenesis
  based on experimental data.  This cannot be achieved by a single
  experiment, but rather a collection of them, based on LHC, ILC,
  neutrino oscillation, neutrinoless double beta decay, direct dark
  matter detection, CMB power spectrum and its tensor mode.
\end{abstract}

\maketitle


\section{Introduction}

Neutrino physics has been full of surprises, and we have learned a lot
in the last ten years.  We would like to learn more.  But what exactly
{\it can}\/ we learn from neutrinos?  Will we ever know the origin of
neutrino mass?  Is it connected to the origin of baryon asymmetry of
the universe?  Possibly the origin of universe itself?

Much of our discussions on this subject have been framed by the famous
seesaw mechanism \cite{seesaw}, the dominant paradigm for the origin
of finite but tiny neutrino mass.  It relies on physics close (but
below) the GUT scale, which is well beyond the reach of any
conceivable accelerator experiments.  Without accessing that energy
scale directly using particle accelerators, how do we ever know if the
seesaw mechanism is true?  Is there a way to test it experimentally?

Unfortunately, the short answer is {\it no}\/.  However, what I would
like to discuss in this talk is that there is a way for us to get
convinced that the seesaw mechanism is right, if Nature is kind to us
\cite{Buckley:2006nv}. 

What we can hope to do is to do a very good job at the accessible
energy scales, namely precision measurements from meV to TeV energies
to fix physics at the low-energy end.  On the other hand, if there is
a way of knowing boundary conditions at high energies, such as the
GUT-scale, we can say something non-trivial about physics between the
two energy scales.

For this program to succeed, we have to be {\it very}\/ lucky, like
all the planets lining up.  But let me remind you that this has gotten
a little easier, now that there are only eight planets to worry about,
not nine.

\section{Why Neutrinos?}

Why have people been interested in neutrinos, especially their mass?
There are good reasons, both from the particle-physics and cosmology
points of view.  

\subsection{Special Role of Neutrino Mass}

It is useful to recall why theorists had always been interested in the
small neutrino masses and their consequences on neutrino oscillation.
It is because we are always interested in probing physics at as high
energies as possible.  One way to probe it is of course to go to the
high-energy collider experiments and study physics at the energy scale
directly.  Another way is to look for rare and/or tiny effects coming
from the high-energy physics.  The neutrino mass belongs to the second
category.

To study rare and/or tiny effects from physics at high energies, we
can always parameterize them in terms of the power series expansion,
\begin{equation}
  {\cal L} = {\cal L}_4 + \frac{1}{\Lambda} {\cal L}_5 +
  \frac{1}{\Lambda^2} {\cal L}_6 + \cdots.
\end{equation}
The zeroth order piece ${\cal L}_4$ is renormalizable and describes the
Standard Model.  On the other hand, the higher order terms are
suppressed by the energy scale of new physics $\Lambda$.  Possible
operators can be classified systematically, which I believe was done
first by Weinberg (but I couldn't find the appropriate reference).
With two powers of suppression, there are many terms one can study:
\begin{equation}
  {\cal L}_6 \supset QQQL,\, \bar{L} \sigma^{\mu\nu} W_{\mu\nu} H e,\,
  {\rm tr}(W_\nu^\mu W_\lambda^\nu W_\mu^\lambda),\,
  \bar{s}d \bar{s}d,\, (H^\dagger D_\mu H) (H^\dagger D^\mu H), \cdots
\end{equation}
The examples here contribute to proton decay, $g-2$, anomalous triple
gauge boson vertex, $K^0$--$\overline{K}^0$ mixing, and the
$\rho$-parameter, respectively.  It is interesting that there is only
one operator suppressed by a single power:
\begin{equation}
  {\cal L}_5 = (LH) (LH).
\end{equation}
After substituting the expectation value of the Higgs, the Lagrangian
becomes
\begin{equation}
  {\cal L} = \frac{1}{\Lambda} (LH)(LH)
  \rightarrow \frac{1}{\Lambda} (L\langle H\rangle)(L\langle H\rangle)
  = m_\nu \nu \nu,
\end{equation}
nothing but the neutrino mass.

Therefore the neutrino mass plays a very unique role.  It is the
lowest-order effect of physics at short distances.  This is a very
tiny effect.  Any kinematical effects of the neutrino mass are
suppressed by $(m_\nu / E_\nu)^2$, and for $m_\nu \sim 0.1$~eV as
suggested by the current data and $E_\nu \sim 1$~GeV for typical
accelerator-based neutrino experiments, it is as small as $(m_\nu /
E_\nu)^2 \sim 10^{-20}$.  At the first sight, there is no hope to
probe such a small number.  However, any physicist knows that
interferometry is a sensitive method to probe extremely tiny effects.
For interferometry to work, we need a coherent source.  Fortunately
there are many coherent sources of neutrinos in Nature, the Sun,
cosmic rays, nuclear reactors (now part of the Nature), etc.  We also
need interference for an interferometer to work.  Because we can't
build half-mirrors for neutrinos, this could have been a show stopper.
Fortunately, there are large mixing angles that make the interference
possible.  We also need long baselines to enhance the tiny effects.
Again fortunately there are many long baselines available, such as the
size of the Sun, the size of the Earth, etc.  Nature was very kind to
provide all necessary conditions for interferometry to us!  Neutrino
interferometry, a.k.a.  neutrino oscillation, is therefore a unique
tool to study physics at very high energy scales.

\begin{figure}
  \centering
  \includegraphics[width=0.5\textwidth]{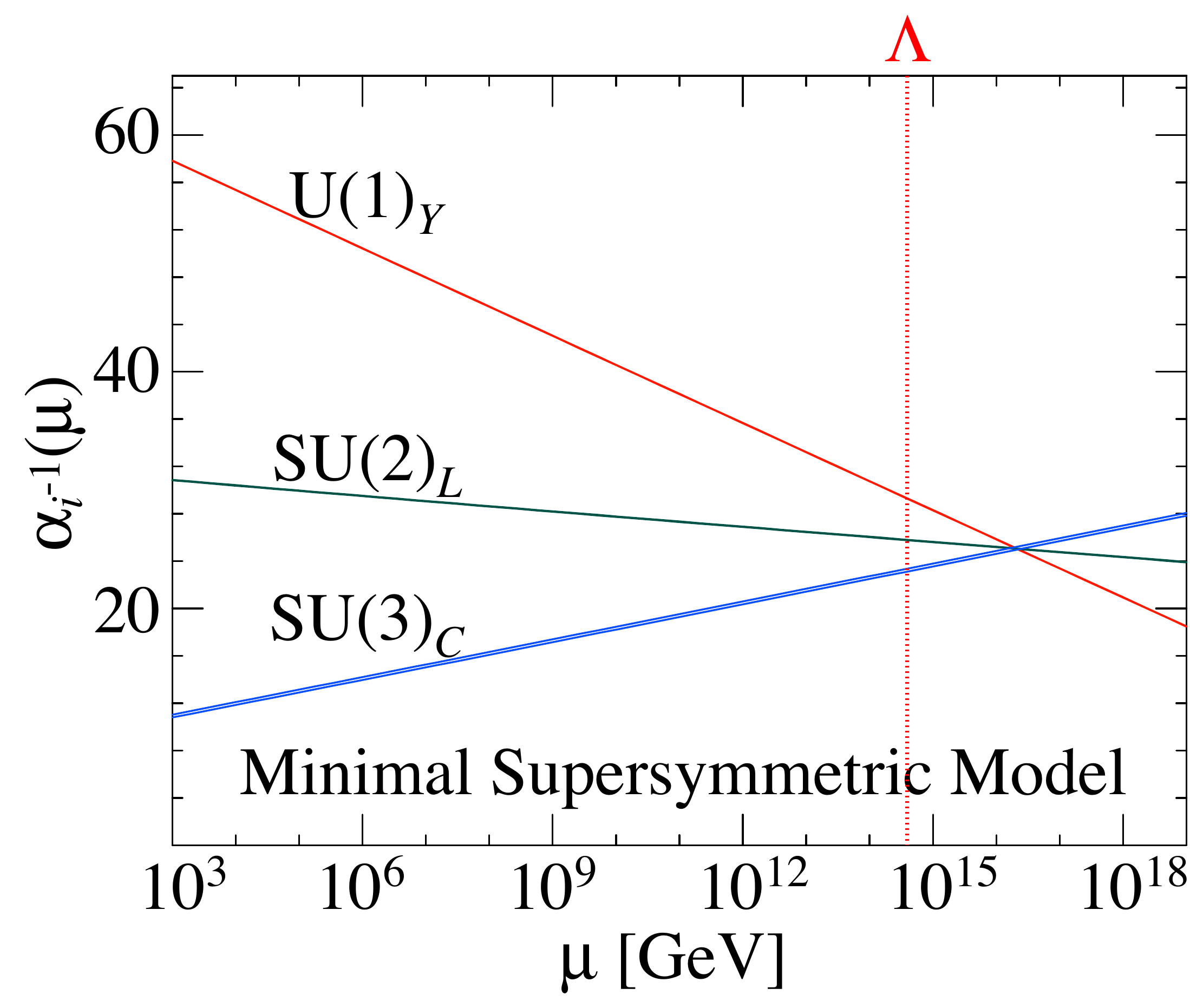}
  \caption{Apparent unification of gauge coupling unification in the
    MSSM at $2 \times 10^{16}$~GeV, compared to the suggested scale of
    new physics from the neutrino oscillation data.}\label{fig:MSSM}
\end{figure}

At the currently accessible energy scale of about a hundred GeV in
accelerators, the electromagnetic, weak, and strong forces have very
different strengths.  But their strengths become the same at $2\times
10^{16}$~GeV if there the Standard Model is extended to become
supersymmetric.  Given this, a natural candidate energy scale for new
physics is $\Lambda \sim 10^{16}$~GeV, which suggests $m_\nu \sim
\langle H \rangle^2/\Lambda \sim 0.003$~eV.  Curiously,
the data suggest numbers
quite close to this expectation.  Therefore neutrino mass under our
current studies may be probing physics at the energy scale of grand
unification.

\subsection{Ubiquitous Neutrinos}

Another reason to be interested in neutrinos is its sheer number in
the universe.  In fact, neutrinos are the most ubiquitous matter
particles in the universe.  They were produced in the Big Bang, when
universe was so dense that neutrinos, despite their only weak
interactions, were in thermal equilibrium with all other particle
species.  Similarly to the cosmic microwave background photons, their
number density had been diluted by the expansion of the universe.  In
comparison, constituents of ordinary matter, electrons, protons, and
neutrons, are far rarer than photons and neutrinos, by about a factor
of ten billions.  It is clear that we need to understand neutrinos in
order to understand our universe.

\begin{figure}[tbp]
  \centering
  \includegraphics[width=0.3\textwidth]{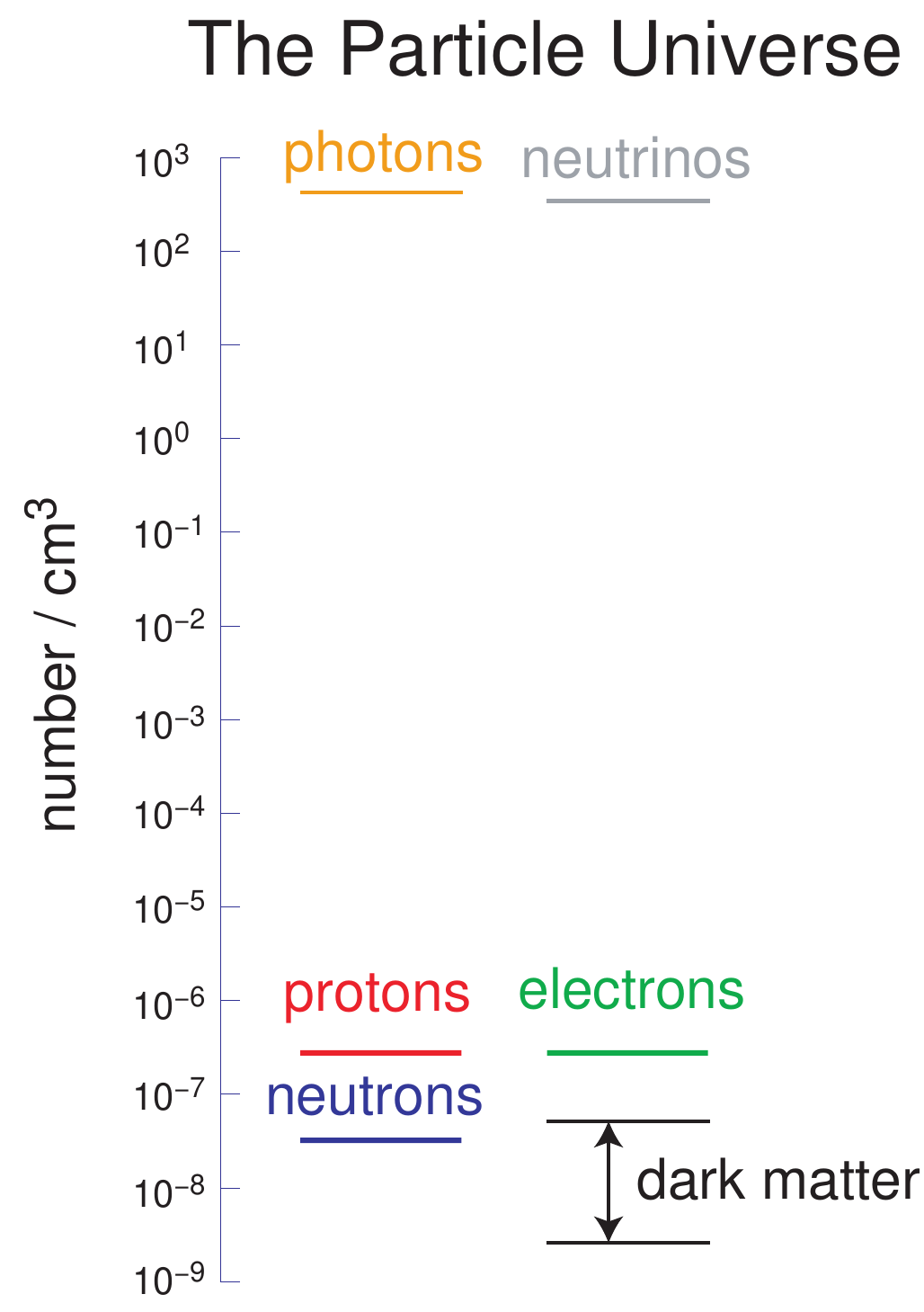}\qquad
  \includegraphics[width=0.3\textwidth]{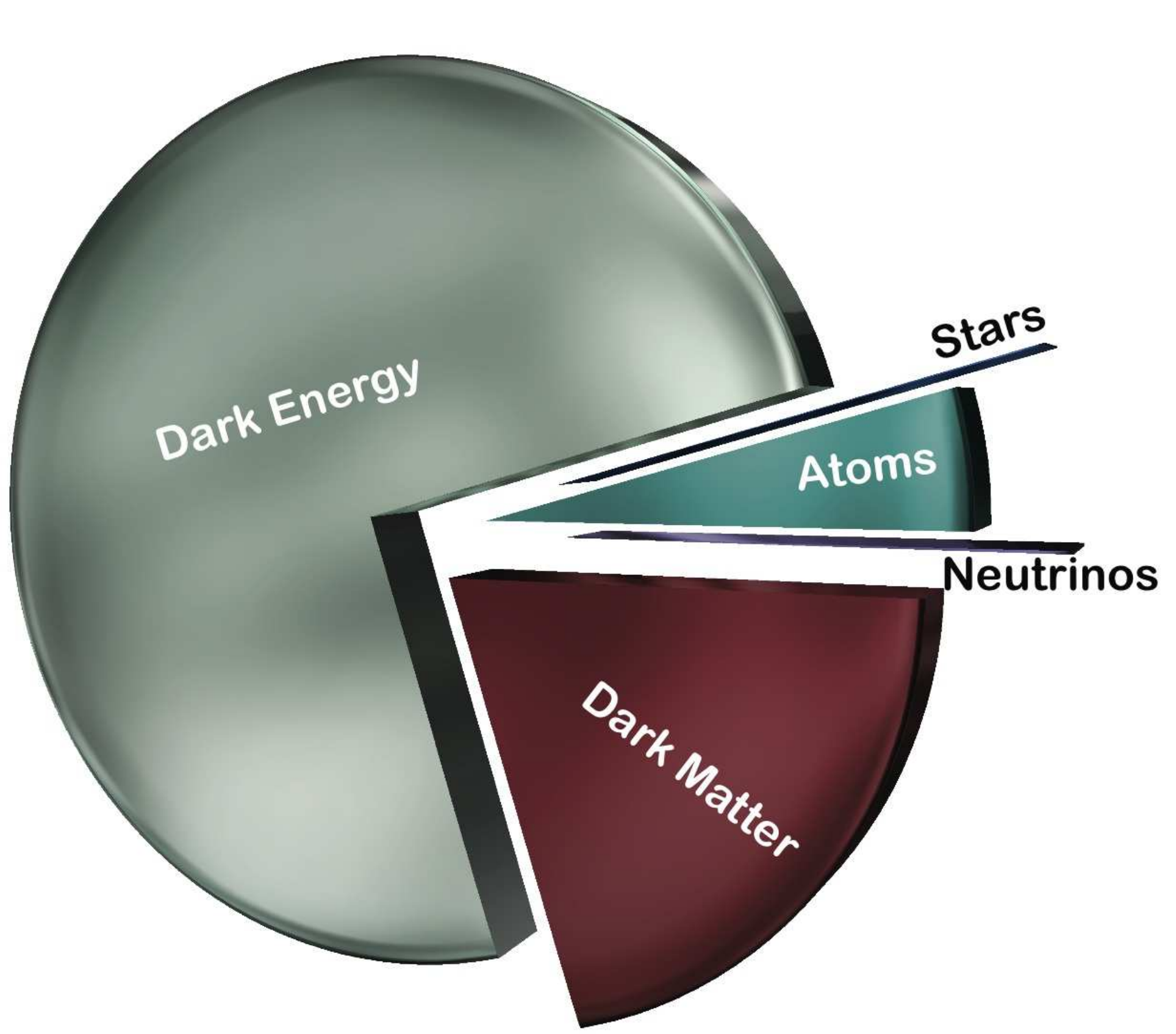}
  \caption{The number density (left) and energy density (right) of
    various components in our universe.}
  \label{fig:universe}
\end{figure}

In terms of energy densities, yet unknown dark matter and dark energy
dominate the universe.  If neutrinos were massless, their energy
density could have been completely negligible in our current
universe.  However, in the last several years, finite mass of
neutrinos had been discovered.  It implies that the neutrinos are as
important as all stars combined.  Unfortunately, we do not know the
mass of neutrinos precisely at this moment, and they may in fact be a
sizable fraction of dark matter.  The precise amount of neutrino
component is relevant to the way galaxies and stars had been formed in
the evolution of the universe.

Neutrinos are important part of the stellar dynamics; without them,
stars would not shine.  There are about $7 \times 10^9 {\rm cm}^{-2}
\sec^{-1}$ neutrinos from the Sun reaching (and streaming through) the
Earth.  They also govern dynamics of supernovae.

\section{What We've Learned}

\subsection{Data}

\begin{figure}
  \centering
  \includegraphics[width=0.5\textwidth]{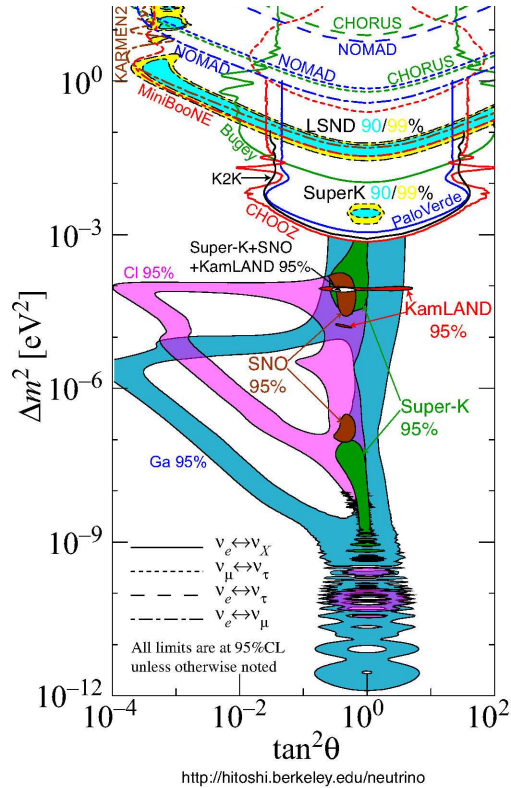}
  \caption{The compilation of the best neutrino oscillation
    data.}\label{fig:nuosc}
\end{figure}

Given the theoretical motivation discussed above, neutrino oscillation
has been searched for over many decades.  Fig.~\ref{fig:nuosc}
summarizes the best information we have on neutrino oscillation.
There are two main parameter regions that are now established by
experimental data.  The oscillation of atmospheric neutrinos is the
first established evidence, which was later confirmed by long-baseline
accelerator neutrino oscillation experiments.  It corresponds to
$\Delta m^2_{23} \approx 2.5 \times 10^{-3}$~eV$^2$ and $\sin^2
\theta_{23} \approx 1$.  The flavor transformation of solar neutrinos
was discovered using water Cherenkov detectors (light and heavy
water), which was established as a consequence of neutrino oscillation
using reactor anti-neutrinos.  It corresponds to $\Delta m^2_{12}
\approx 7.6 \times 10^{-5}$~eV$^2$ and $\tan^2 \theta_{12} \approx 0.5$.

In both cases, the mixing angle came out large, which allowed us to
observe neutrino oscillation.  This is the generosity of Nature we
discussed.  At the same time, the neutrino mass scale is {\it
  tiny}\/.  According to the previous discussion, it is a good
indication that we are probing physics at {\it very}\/ high energy
scales.

\subsection{Surprises}

It is useful to recall what a typical theorist used to say back around
1990.  
\begin{itemize}
\item The solution to the solar neutrino problem must be the small
  mixing angle MSW solution because it is so beautiful.
\item The natural scale for $\nu_\mu \rightarrow \nu_\tau$ oscillation
  is $\Delta m^2 \sim {\rm eV}^2$ because it is the cosmologically
  interesting range.
\item The angle $\theta_{23}$ must be of the same order of magnitude
  as $V_{cb}$ because of the grand unification.
\item The atmospheric neutrino anomaly must go away because it would
  require a large mixing angle to explain.
\end{itemize}
Needless to say, theorists have a {\it very good track record}\/ in
neutrino physics.

Indeed, the recent results from neutrino oscillation physics had
surprised almost everybody.  The prejudice has been that the mixing
angles must be small because quark mixing angles are small, and the
masses must be hierarchical because both quarks and charged lepton
masses are hierarchical.  Given that the LMA is now chosen, all mixing
angles are large except for $U_{e3}$ that must be small-ish (but the
current limit is not very strong, $|U_{e3}| \lesssim 0.2$).

Another surprise is their mass spectrum.  The quarks and charged
leptons have so-called hierarchical mass spectra, namely that the
masses of similar types are drastically different among three
generations of elementary particles.  For instance, the masses of the
up- and top-quarks are different by five orders of magnitude.  On the
other hand, two heavier neutrino masses differ {\it at most}\/ by a
factor of about five, possibly even degenerate.  

However, the measurements of neutrino masses and mixings are still
incomplete.  Future measurements are likely to bring more surprises.

\begin{figure}
  \centering
  \includegraphics[width=0.5\textwidth]{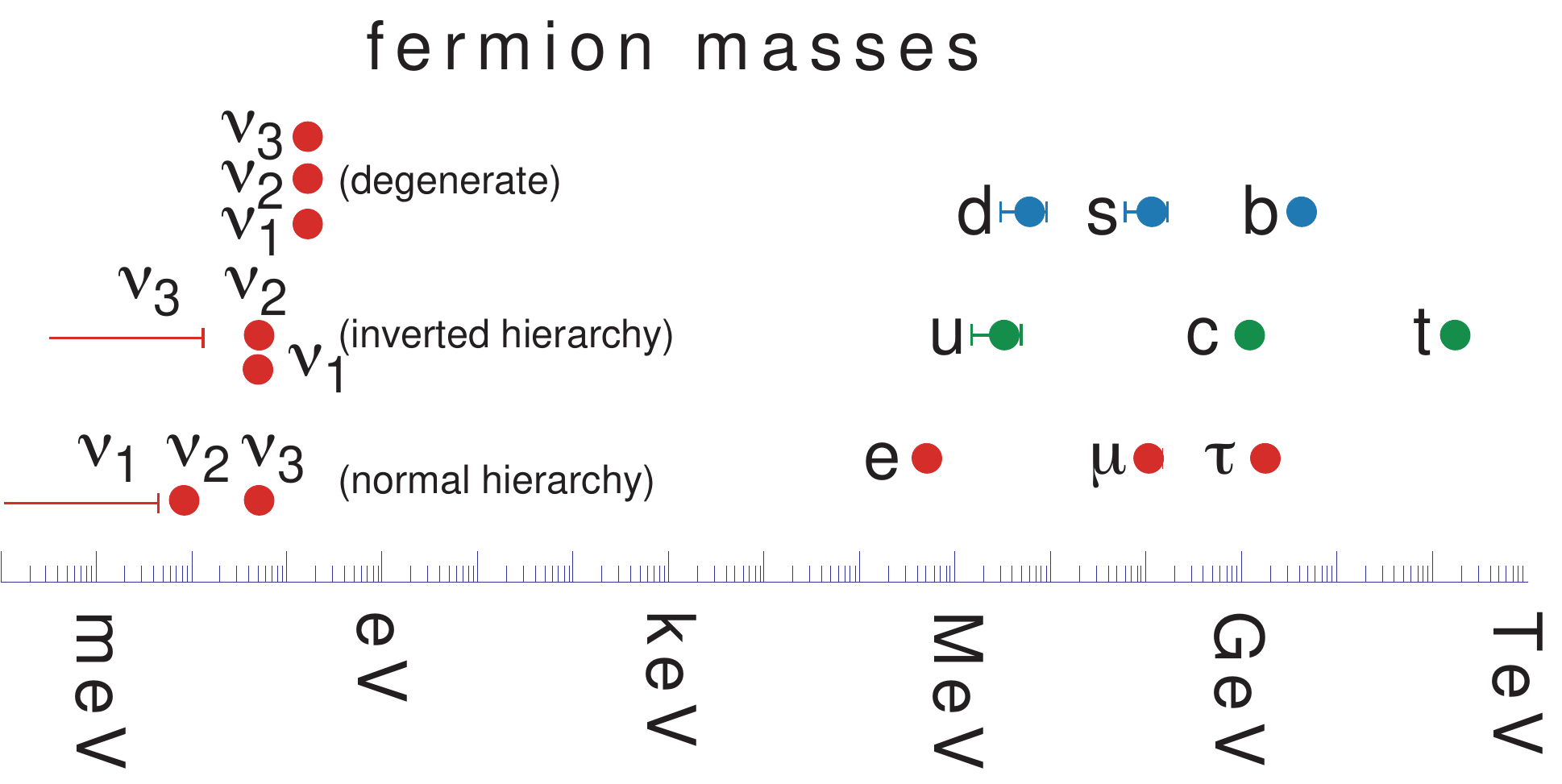}
  \caption{Mass spectrum of known elementary fermions.  Neutrino
    masses are not well-known and are meant to be only schematic for
    three possibilities, degenerate, inverted hierarchy, or normal
    hierarchy.}\label{fig:masses}
\end{figure}

\subsection{The Big Questions}

The discussions above lead to the following set of big questions
concerning neutrinos:
\begin{itemize}
\item What is the origin of neutrino mass?
\item Did neutrinos play a role in our existence?
\item Did neutrinos play a role in forming galaxies?
\item Did neutrinos play a role in the birth of the universe?
\item Are neutrinos telling us something about unification of matter
  and/or forces?
\end{itemize}
These are {\it big}\/ questions, and by definition are very difficult
questions to answer.

What we try to see is if future data will address any of these big
questions.  Clearly no guarantee.  But I'd like to argue that there is
a chance.  That is basically the whole point of my talk here.

\section{Seesaw and SUSY-GUT}

The famous seesaw mechanism is meant to be an explanation why neutrino
mass is finite but yet tiny.  You introduce right-handed neutrinos,
fermion species completely neutral under the gauge groups of the
standard model.

The Lagrangian (for the case of single flavor) is extremely simple.
The right-handed neutrino $N$ has the Yukawa coupling to Higgs boson
the same way as any other fermion species we know (charged leptons and
quarks), while it is neutral under the standard model gauge groups and
is allowed to have a Majorana mass term
\begin{equation}
  {\cal L} = - y \bar{L} N H - \frac{1}{2} M N N + {\it c.c.}
\end{equation}
Upon the condensation of the Higgs boson in the universe $\langle H
\rangle = v$, we find a mass matrix of neutrinos,
\begin{equation}
  \frac{1}{2} (\nu, N) \left(
    \begin{array}{cc}
      0 & y v\\ y v & M
    \end{array} \right) \left(
    \begin{array}{cc}
      \nu\\ N
    \end{array} \right).
\end{equation}
Because the Majorana mass $M$ does not rely on the electroweak
symmetry breaking it may well be much higher, $M \gg v$.  Then one of
the eigenvalues of this mass matrix would be simply $M$.  On the other
hand, due to the vanishing (1,1) entry, the product of two
eigenvalues, namely the determinant, is $- (yv)^2$.  Therefore the
other smaller eigenvalue must be $-(yv)^2/M$, which is much smaller
than the electroweak scale if $M \gg v$.  This way, we can understand
why the neutrino mass is much lighter than other quark and charged
lepton masses of $O(yv)$ because of the heavy right-handed neutrino
mass scale $M$.  The heavier $M$ is, the lighter the neutrino mass
becomes, hence the seesaw.

If we naively assume $y \approx 1$ (as suggested by the top Yukawa
coupling) and take $m_\nu \approx (\Delta m^2_{23})^{1/2} \approx
0.05$~eV, we find $M \approx \mbox{a few} \times 10^{14}$~GeV.

Indeed, there is a reason to think that the seesaw scale is related to
the grand unified theory.  For instance, there is a special interest
in the unified group SO(10).  It is the smallest gauge group which has
complex representations yet is automatically anomaly-free.  Its
smallest representation (16) accommodates all of the observed fermions
{\it and}\/ right-handed neutrinos.  If this is the case, the Majorana
mass of right-handed neutrinos is not allowed above the GUT-scale
because of its complex nature, while can be induced by the breaking of
the GUT group down to the standard model.  This way, the seesaw scale
$M$ is tied to the GUT-scale as $M \approx M_{GUT}$, if SO(10) is
broken by 126-Higgs, while $M \approx M_{GUT}^2/M_{Planck}$ if broken
by 16-Higgs.  The latter appears to be a good numerical match to the
neutrino mass seen in experiments.

In addition to inducing finite but tiny neutrino mass, right-handed
neutrinos may have played a critical role for our existence in the
universe.  Given that the neutrinos come in three families (actually
two is enough for the purpose below), they may violate CP.  The
right-handed neutrinos are expected to be heavy, but the early
universe may have produced them at temperatures above their masses.
At the tree-level, they decay 50:50 into leptons and anti-leptons.
However the interference between the absorptive part of the one-loop
diagram and the tree-level amplitude can violate CP and makes the
branching fractions a little different.  Namely that the right-handed
neutrinos may decay preferentially into anti-leptons over leptons,
creating a (negative) {\it lepton asymmetry}\/ of the universe.

It turns out that this created lepton asymmetry gets partially
converted to the baryon asymmetry by the standard model gauge
interaction.  This is because both the lepton and baryon number
currents are not conserved in the $SU(2)$ gauge theory because of its
chiral nature,
\begin{equation}
  \partial_\mu j^\mu_B = \partial_\mu j^\mu_L 
  \propto \epsilon^{\mu\nu\rho\sigma} {\rm tr} W_{\mu\nu} W_{\rho\sigma}
  \neq 0.
\end{equation}
It allows for the baryon number and lepton number to change by an
equal amount due to the thermal fluctuations of the $W$-fields.  In
other words, the (negative) lepton asymmetry can be converted to the
(positive) baryon asymmetry.  The question then is if this conversion
actually occurs.  Simply by thermodynamic considerations, keeping the
asymmetry only in leptons costs more free energy than distributing the
asymmetry both among leptons and quarks.  Therefore, the lepton
asymmetry {\it does}\/ get partially converted to the baryon
asymmetry.  This is what is called leptogenesis \cite{leptogenesis}.

Right-handed neutrinos might have played an even bigger role; they may
be the origin of the universe \cite{inflation}.  The idea of grand
unification has the hierarchy problem, and the standard model particle
content is actually not consistent with the observed values of the
gauge coupling constants.  Supersymmetry solves both problems.  Once
we have supersymmetry, the right-handed neutrinos come with their
scalar partners.  Because of their large mass, the potential for the
right-handed sneutrinos is dominated by their mass term, $M^2
|\tilde{N}|^2$.  Interestingly, this simple potential is what may make
the cosmological inflation possible.  The universe starts out
microscopically small with an amplitude of right-handed sneutrino
larger than the Planck scale.  Then the right-handed sneutrino rolls
down the potential slowly, thereby expanding the universe to a
macroscopic size.  Eventually the amplitude becomes less than the
Planck scale and the exponential expansion stops.  The right-handed
sneutrino oscillates around the minimum of the potential (zero) and
behaves as matter, which is nothing but a Bose--Einstein condensate of
the right-handed sneutrino.  It then decays, just like the decay of
right-handed neutrinos discussed above, which can create a negative
lepton asymmetry.  Namely that the right-handed sneutrinos may play
the role of the origin of the universe as well as our existence.

This is all interesting, and a large amount of ink has been devoted to
further elaboration of these subjects.  Now we come to the central
question in my talk.

\section{Experimental Test}

As we have seen above, the seesaw mechanism is quite attractive,
especially in conjunction with the GUT, leptogenesis, or inflation.
Is there a way of proving this mechanism?

Obviously the short answer is {\it no}\/.  Here is an irony: the
reason why the seesaw mechanism is so attractive is because it refers
to physics well beyond the reach of any conceivable particle
accelerators, while this very point makes it impossible to prove.  Is
it hopeless?

I'd like to argue that the future experiments may provide us
information which will make us believe that the seesaw mechanism is
right after all \cite{Buckley:2006nv}.  It is not a direct proof as we
became accustomed to in particle physics, such as the precision
electroweak measurements that completely convince us of the validity
of the gauge theory.  It will rather be a collection of circumstantial
evidences which so strongly argue for it that we will buy into
it.\footnote{In other words, we may not convict ``O.J. Simpson'' in a
  criminal court, but possibly in a civil case.}

This is not to be ashamed of.  Observational cosmology has made an
enormous progress in recent years, but it is and will be a form of
archaeology.  We never redo the Big Bang.\footnote{Boris Kayser once
  remarked that a severe funding problem prevents us from attempting
  it.}  Nonetheless that power spectrum in the cosmic microwave
anisotropy is so precisely measured and theoretically clean that we
are (most of the time) happy to accept its implications, such as the
energy budget, age of the universe, non-baryonic dark matter, etc.
This is the best we can hope for in the case of the seesaw mechanism.

\subsection{Proving Seesaw}

One such example was worked out by my student Matt Buckley and myself
\cite{Buckley:2006nv}.  If we will see the following outcomes from the
future experiments, we will all (except for some die-hard skeptics)
believe in the seesaw mechanism:
\begin{itemize}
\item Discovery of Supersymmetry at LHC and ILC, whose precision
  measurements confirm its mass spectrum as predicted by GUT.
\item Discovery of neutrinoless double beta decay.
\end{itemize}
In addition, other supporting measurements will boost the credibility
of the conclusion:
\begin{itemize}
\item Consistency of the SUSY parameters measured at colliders and the
  cosmological abundance of the Lightest Supersymmetric Particle.
\item Improved limits or discovery of Lepton Flavor Violation.
\end{itemize}
I will explain why this is so in the remainder of this talk.

If supersymmetry is discovered, the precision measurement of its mass
spectrum is extremely important \cite{Tsukamoto:1993gt}.  The case of
our interest here is if the masses {\it unify}\/.

As we discussed earlier, the observed gauge coupling constants are
consistent with the GUT assuming the particle content of the minimal
supersymmetric standard model.  We know that the unification scale is
approximately $2 \times 10^{16}$~GeV.  But it is basically three
straight lines meeting at a point.  One may say that two lines always
meet, with 50:50 chance meeting at an energy scale above (not below)
TeV.  Then it may be just a numerical coincidence that the third line
appears to go through the same point.

Once we measure the gaugino masses, the situation will be quite
different.  There are three gaugino masses, $M_1$ (bino), $M_2$
(wino), $M_3$ (gluino).  If two of them meet at the {\it same energy
  scale}\/ where the gauge coupling constants appear to meet, this is
the second coincidence.  If the third gaugino mass meets at the same
energy and mass, it is the third.  The case for unification becomes
far stronger.

Then come the masses of sfermions, namely the superpartners of quarks
and leptons.  For each generation in a GUT, left-handed quark doublet
(in three colors=six degrees of freedom), right-handed up quark (in
three colors), and right-handed charged lepton are unified in a
decouplet ($3\times 2+3+1=10$), while left-handed lepton doublet
(charged lepton and neutrino) and right-handed down quark (in three
colors) in a quintet ($2+3=5$).  The masses of their superpartners
therefore unify at the same energy scale where both gauge coupling
constants and gaugino masses unify.  This gives three more
coincidences {\it per generation}\/.

Once the data show the gaugino mass and sfermion mass unification, it
could still be just an accident, but the coincidences comes in such a
multitude that we will be led to believe there is indeed unification.
Probably some die-hard skeptics will remain, but the GUT becomes
extremely compelling and undeniable.

Such an experimental observation would tell us several things.  First
of all, there is unification at $2\times 10^{16}$~GeV.  Second, given
its successful description in terms of a simple gauge theory, the
renormalizable quantum field theory must be applicable up to this
energy scale.  Third, we can then use the GUT as the boundary
condition for physics at high energies, so that we can start
constraining physics below the GUT scale.

On the other hand, an observation of neutrinoless double beta decay
would tell us that there is indeed the Majorana neutrino mass operator
\begin{equation}
  {\cal L}_5 = \frac{1}{\Lambda} (LH)(LH),
\end{equation}
where the scale $\Lambda$ is significantly below the GUT scale.  Since
this operator is non-renormalizable, it needs to be generated by
integrating out some heavy particles below the GUT scale.  In other
words, there is ``new physics'' below the GUT scale responsible for
generating this operator, and therefore we need to add new particles
beyond the minimal supersymmetric standard model.

\begin{figure}
  \centering
  \includegraphics[width=0.5\textwidth]{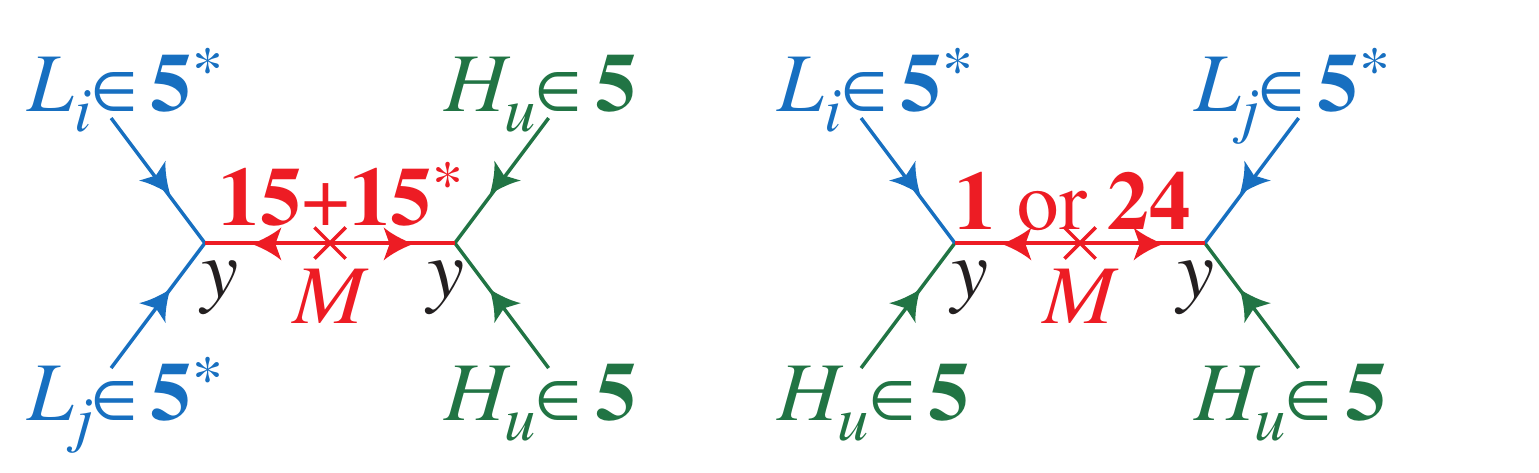}
  \caption{Three possible sets of new particles below the GUT-scale
    that would preserve the unification of gauge coupling constants
    and gaugino masses, and can generate Majorana neutrino
    mass.}\label{fig:seesaw}
\end{figure}

What could such new particles be?  Given the non-trivial success of
unification, new particles should not spoil it.  It is well-known that
the addition of complete $SU(5)$ multiplets would maintain the
unification of gauge coupling constants.  It turns out that complete
multiplets would maintain the unification of gaugino masses as well
\cite{Kawamura:1993uf}.  There are three possible $SU(5)$ multiplets
that can generate the Majorana neutrino mass operator for all three
generations.\footnote{It is possible that only two out of three
  neutrinos have mass as far as the current data are concerned.  We
  assume here that all three have mass, but the rest of the
  discussions can be trivially modified if that is not the case.}  We
add either three adjoints (24), three singlets (1), or one symmetric
tensor ($15+\overline{15}$).  The last possibility is often called
Type-II seesaw, while the standard (Type-I) seesaw we discussed
earlier is the second possibility.

\begin{figure}
  \centering
  \includegraphics[width=0.45\textwidth]{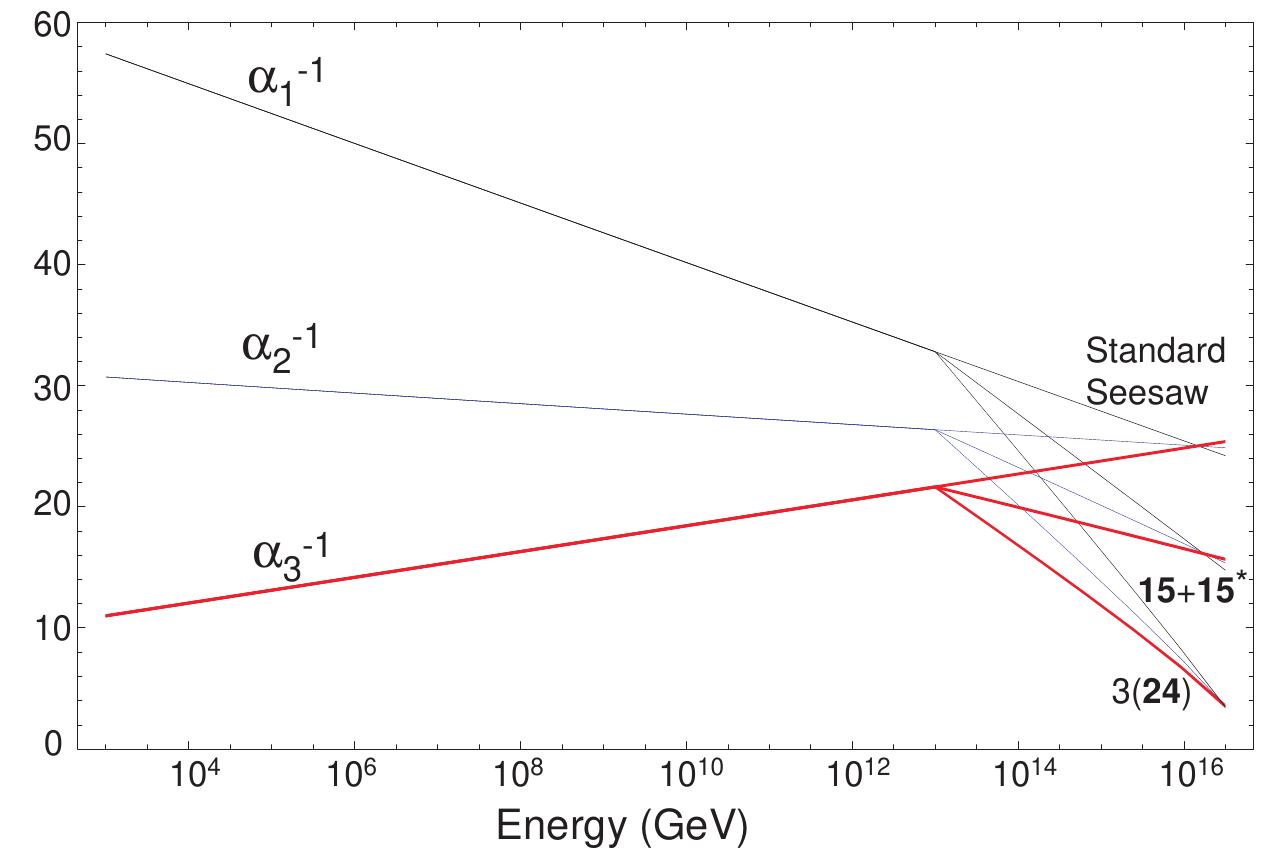}
  \includegraphics[width=0.45\textwidth]{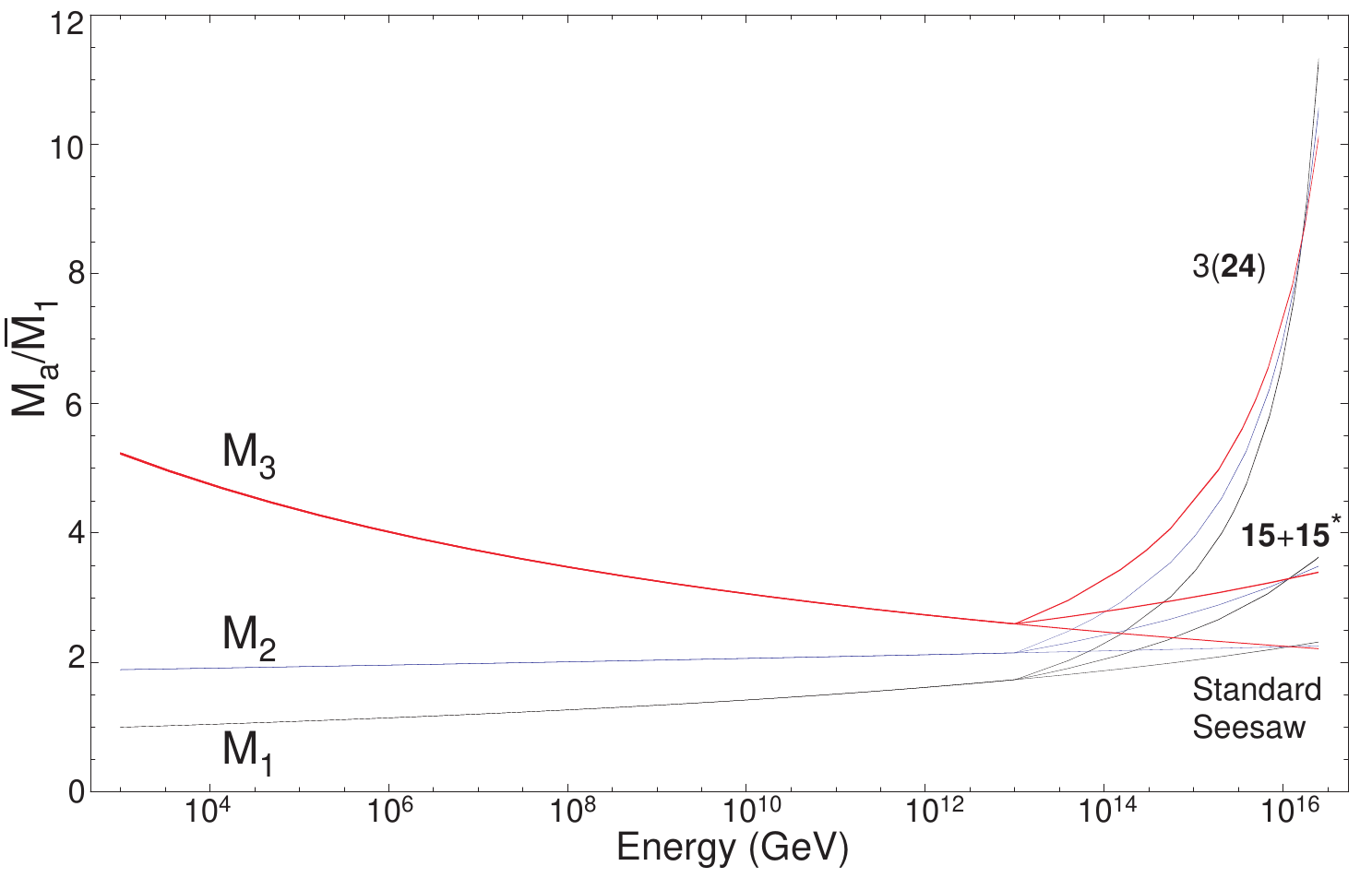}
  \caption{From the low-energy point of view, the new sets of
    particles cannot be discerned based on the unification of gauge
    coupling constants and gaugino masses.}
\end{figure}

However, the scalar mass unification is affected
\cite{Kawamura:1993uf}.  Depending on the three options, the observed
spectrum may or many not unify at the GUT scale.  This is because the
gauge-non-singlet particles affect the running of gauge coupling
constants and gaugino masses, which in turn feed into the scalar
masses-squared.  Having seen the data consistent with simple
unification, the gauge non-singlet options will be excluded;
indirectly establishing the gauge singlet options for the origin of
the neutrino mass, namely the seesaw mechanism.

\begin{figure}
  \centering
  \includegraphics[width=0.5\textwidth]{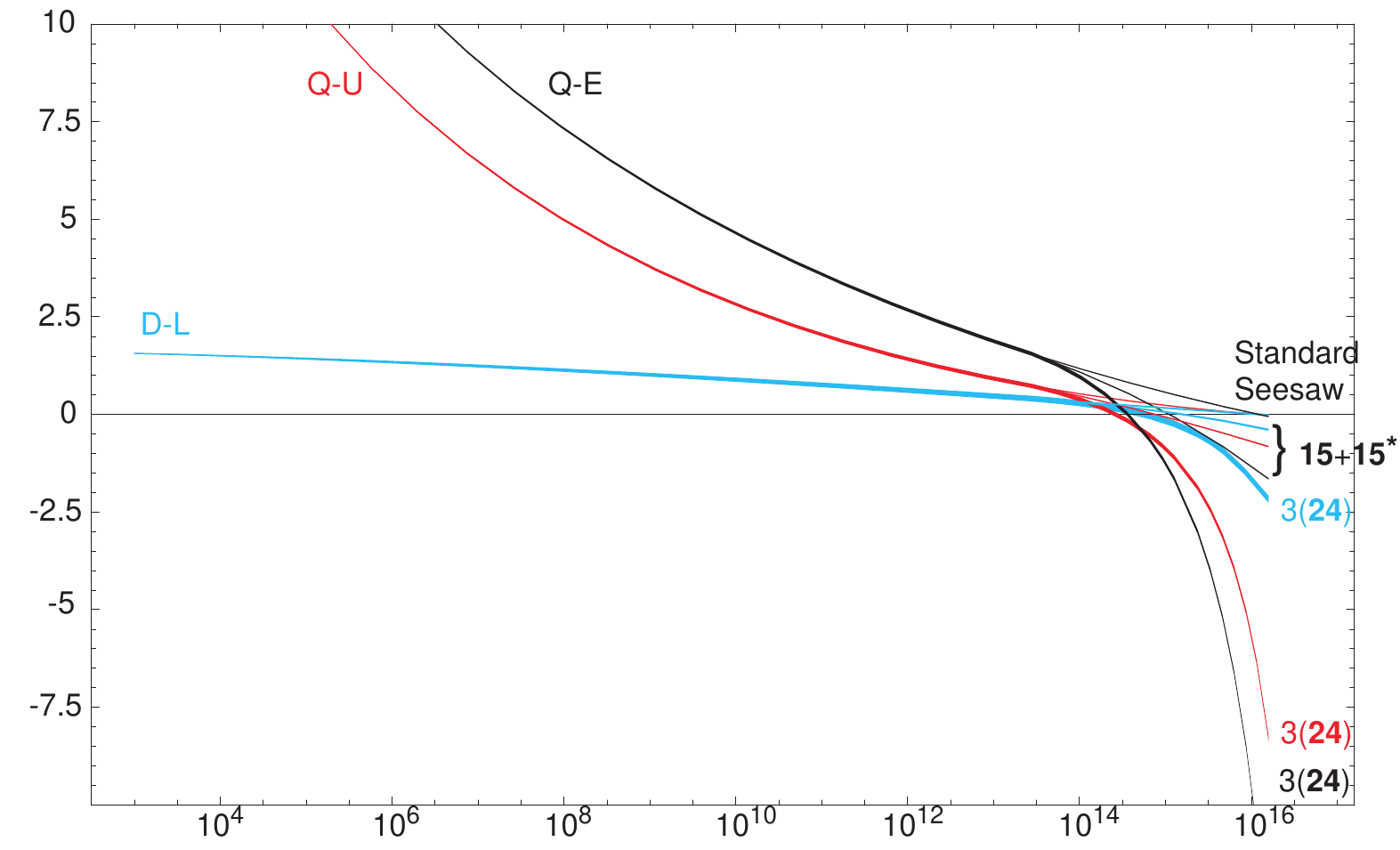}
  \caption{The possibilities that cannot be discerned based on the
    unification of gauge coupling constants and gaugino masses can be
    differentiated using the sfermion masses.}\label{fig:bottomuppt1}
\end{figure}

This way, we may well be convinced that the seesaw mechanism is the
origin of the neutrino mass.

There are a few loose ends, which can be tied up by additional data.
For instance, once both supersymmetry and neutrinoless double beta
decay are discovered, one may suspect that the neutrinoless double
beta decay may be due to the $R$-parity violation, not Majorana
neutrino mass.  This ambiguity can be resolved if the following
happens.  Once the supersymmetry parameters are measured accurately,
we will be able to compute the abundance of the Lightest
Supersymmetric Particle (LSP) \cite{Baltz:2006fm}.  If it agrees with
the cosmological abundance of dark matter, it can be taken as a very
strong evidence that the dark matter of the universe is indeed the
LSP.  Then the LSP must live much longer than the age of the universe,
and hence the $R$-parity must be a very good symmetry.  In addition,
the scattering cross section of LSP on nucleus can also be compared to
data if direct detection experiments succeed to detect dark matter.
Overall, we may well know that the neutrino mass is Majorana.

\begin{figure}
  \centering
  \includegraphics[width=0.3\textwidth]{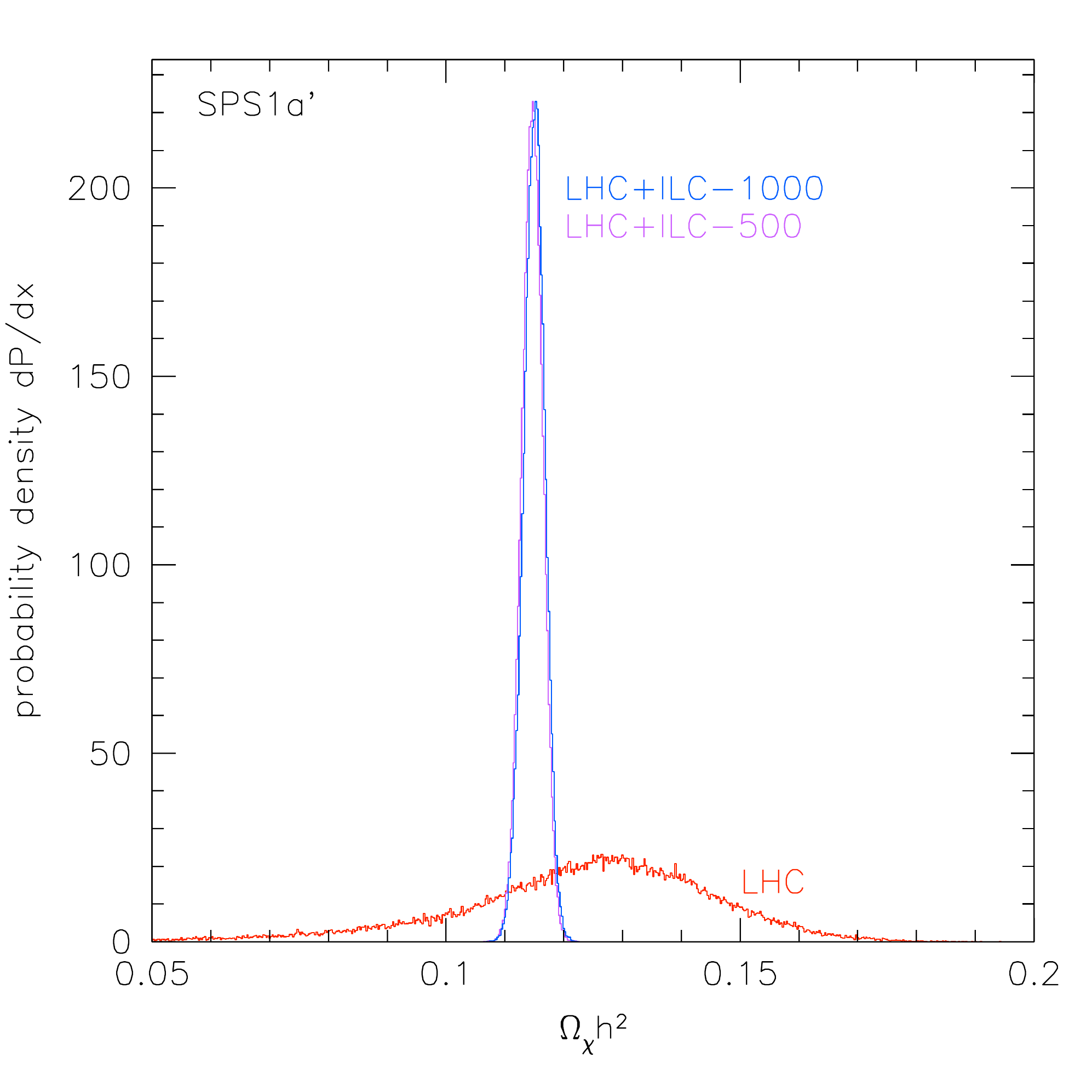}
  \caption{The consistency check between the collider data and dark
    matter abundance \cite{Baltz:2006fm}.}\label{fig:lcc5_oh2}
\end{figure}

Another possible loose end is the mass scale $M$ of the additional
particles needed to generate the Majorana neutrino mass operator.  If
their masses are very close to the GUT scale, their impact on the
running of fermion masses will be small.  One way this can be avoided
is if the neutrinoless double beta decay suggests rather high mass of
neutrinos $\sim 0.1$~eV.  Another way is to know that Lepton Flavor
Violation (LFV) is suppressed.  The higher their masses, the larger
the Yukawa couplings $y$ must be to reproduce the same neutrino mass
$\propto y^2/M$.  Then their loops introduce flavor-dependent effects
in the slepton masses-squared $\propto y^2$, inducing LFV at low
energies.  Therefore small LFV suggests low $M$, and hence a certain
minimum size in the running of fermion masses-squared.

\begin{figure}
  \centering
  \includegraphics[width=0.5\textwidth]{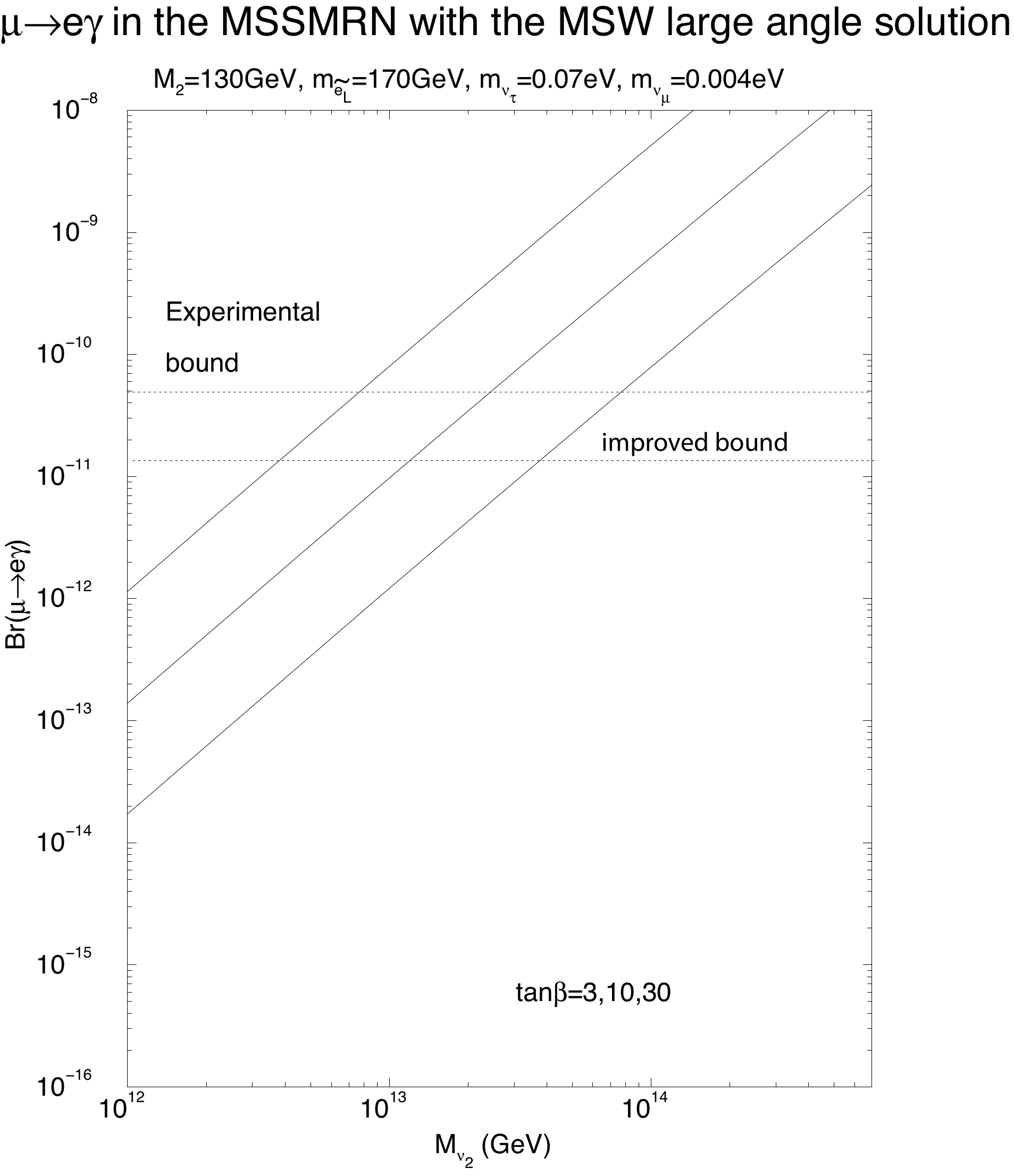}
  \caption{The $\mu \rightarrow e\gamma$ branching fraction due to the
    loop effects of sleptons.  The seesaw mass scale enters the
    prediction because of the loops due to the right-handed neutrinos
    in the slepton masses-squared.  An updated experimental limit is
    superimposed since the original paper
    \cite{Hisano:1998fj}.}\label{fig:MSSMmeMl}
\end{figure}

\subsection{Leptogenesis}

OK, it may be possible to prove the seesaw mechanism if the SUSY-GUT
turns out to be correct as we've discussed.  What about leptogenesis?

That is obviously much harder.  Certainly, we may find additional
circumstantial evidence.  We may find $\theta_{13}$ is not too small
which leptogenesis models tend to prefer.  We may find CP violation in
neutrino oscillation, establishing that CP is violated in the lepton
sector.  But both of them are plausibility tests, not really an
experimental evidence.

We can certainly do better.  Given the gravitino problem
\cite{Kawasaki:2004qu}, we know that the reheating temperature after
inflation must be low, which in turn requires whatever particles
responsible for baryogenesis must be much lighter than the GUT-scale.
On the other hand, once the superparticle spectrum is consistent with
simple SUSY-GUTs, we know there are no extra particles charged under
the standard model gauge groups well below the GUT-scale, as we
discussed above.  Then it leaves only two options: baryogenesis with
gauge singlets, or use particles in the Minimal Supersymmetric
Standard Model.

The latter possibility implies the electroweak baryogensis.  We know
it is already tightly constrained by the search for electric dipole
moments (electron, neutron, Mercury) and the Higgs search.  (The
latter constraint may be relaxed if the Higgs sector is more
complicated, such as the Next-to-Minimal Supersymmetric Standard
Model.)  Nonetheless we need light bosons to make the electroweak
phase transition first order, such as stop, which may be excluded by
collider searches.  Overall, we may well exclude the electroweak
baryogenesis using a combination of collider and EDM searches.  Then
it only leaves the option of baryogenesis via gauge singlets.  It
awfully looks like leptogenesis.  It is certainly not a proof, but
very compelling.\footnote{I do not know at this moment if
  Affleck--Dine baryogenesis is a way out of this argument.}

If we are lucky, we may do even better.  The improved cosmological
data may find the spectral index and the tensor component consistent
with the $\phi^2$ chaotic inflation, $n_s \approx 0.96$, $r \approx
0.16$, with $M \approx 10^{13}$~GeV.  Since this mass is well below
the GUT-scale, it has to be a gauge singlet.  Here, the search for the
$B$-mode fluctuation in the cosmic microwave background anisotropy
(Planck, Polarbear, Quiet, etc) is crucial.  But having established
that the seesaw mechanism is correct, it is tempting to identify the
inflaton $\phi$ with the right-handed neutrino which we know should
exist \cite{inflation}.  Then the reheating process itself is
leptogenesis.

\begin{figure}
  \centering
  \includegraphics[width=0.3\textwidth]{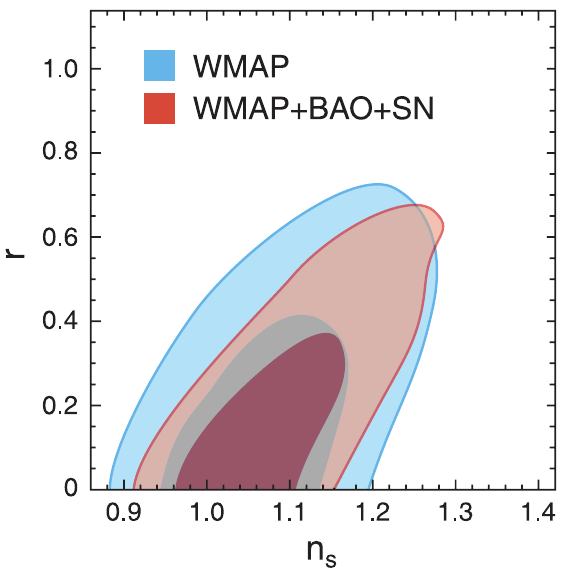}
  \caption{The current limit on $(n_s, r)$ \cite{Komatsu:2008hk}.  The
    $\phi^2$ chaotic inflation is still consistent with the
    data.}\label{fig:f4}
\end{figure}

\section{Conclusions}

We are going through a revolution in neutrino physics, and are excited
that the discovered neutrino mass may tell us something about physics
at very high-energy scales.  Unfortunately this very nature of
neutrino mass makes it impossible to directly access its origin.
However, by a collection of experiments, the seesaw mechanism may be
revealed as the only consistent explanation for the origin of neutrino
mass.  We may further find circumstantial evidence that no other
mechanism of baryogenesis, other than leptogenesis, works.  If we are
lucky, we may even find a compelling reason that the seesaw mechanism
is the origin of the universe, namely that the right-handed neutrino
is an inflaton.

Planets do not often line up, but they sometimes do.  We just hope
that Nature is as kind as she has been to us.

\begin{theacknowledgments}
  This talk is primarily based on a work with my student Matt Buckley,
  but also on a collection of works with many collaborators including
  Keisuke Fujii, Toshifumi Tsukamoto, Masahiro Yamaguchi, Yasuhiro
  Okada, Yoshiharu Kawamura.  Ippp thank them for fruitful discussions
  and collaborations.  This work was supported in part by World
  Premier International Research Center Initiative (WPI Program),
  MEXT, Japan, in part by the U.S. DOE under Contract
  DE-AC03-76SF00098, and in part by the NSF under grant PHY-04-57315.
\end{theacknowledgments}

\end{document}